\documentclass[aps,pra,preprintnumbers,twocolumn,floatfix]{revtex4}

\usepackage{hyperref}
\usepackage{amssymb}
\usepackage{color}
\usepackage{graphicx}
\usepackage{subfigure}
\usepackage{epsfig}
\usepackage{amsmath}
\usepackage{subfigure}
\usepackage[normalem]{ulem}

\begin{document}

\title{Cut-and-paste restoration of entanglement transmission}
 
\author{\'{A}lvaro Cuevas$^1$, Andrea Mari$^2$, Antonella De Pasquale$^2$, Adeline Orieux$^{1,3}$, Marcello Massaro$^{1,4}$, Fabio Sciarrino$^1$, Paolo Mataloni$^1$, and
Vittorio Giovannetti$^2$}
\affiliation{$^1$Universit\`{a} di Roma La Sapienza, Dipartimento di Fisica, Rome, Italy,\\
$^2$NEST, Scuola Normale Superiore and Istituto Nanoscienze-CNR, I-56127 Pisa, Italy,\\
$^3$LTCI, CNRS, T\'el\'ecom ParisTech, Universit\'e Paris-Saclay, 75013, Paris, France\\
$^4$Integrated Quantum Optics Group, Applied Physics, University of Paderborn, 33098 Paderborn, Germany.}

\begin{abstract}
The distribution of entangled quantum systems among two or more nodes of a network is a key task at the basis of quantum communication, quantum computation and quantum cryptography. 
Unfortunately the transmission lines  used in this procedure can introduce so much perturbations and noise in the transmitted signal that prevent the possibility 
of restoring quantum correlations in the received messages either by means of encoding optimization or by exploiting local operations and classical communication.
In this work we present a procedure which allows one to improve the performance of some of these channels.
 The mechanism underpinning this result is a protocol which we dub {\it cut-and-paste}, as it consists in extracting and reshuffling the sub-components of these communication lines, which finally succeed in ``correcting each other". The proof of this counterintuitive phenomenon, while improving our theoretical understanding of  quantum entanglement, has also a direct application in the realization of quantum information networks based on imperfect and highly noisy communication lines.
A quantum optics experiment,  based on the transmission of single-photon polarization states, is also presented which provides a proof-of-principle test of the proposed protocol. 
\end{abstract}

\maketitle

\section{Introduction}

Reliable quantum communication channels \cite{gordon1962,holevo2012} are crucial for all quantum information and computation protocols \cite{nielsen2010} where quantum data  must be faithfully and efficiently transmitted among different nodes of a network \cite{kimble2008}, usually via traveling optical photons \cite{duan2001, caves1994}. Typical applications are quantum teleportation \cite{bennet1993}, cryptography \cite{gisin2002} and distributed quantum computation \cite{serafini2006}.

According to quantum mechanics, two systems can be prepared in an entangled state characterized by extraordinary correlations that are beyond any  classical description \cite{horodecki2009}. A key objective in quantum communication is the distribution of such entanglement between two parties, say Alice and Bob. This can be easily achieved with perfect quantum channels: Alice entangles two systems and sends one of them to Bob. Even if the channel is not perfect, a  fraction of the initial entanglement can still survive the transmission process and may be successively amplified by distillation protocols \cite{bennet1996}. In principle, this technique allows for efficient entanglement distribution along a network of  quantum repeaters connected by imperfect channels \cite{briegel1998,sangouard2011}. 

The situation is completely different for {\it entanglement-breaking} (EB) channels \cite{horodecki2003} which 
  are so leaky and noisy that entanglement is always destroyed for every choice of the initial state. Such communication lines  behave essentially as classical {\it measure-and-reprepear} operations \cite{holevo1999}. In this case the previous techniques based on entanglement distillation and quantum repeaters cannot be applied simply because, after the transmission process, there is nothing left to distill or to amplify. Furthermore, any other error-correction protocol, based on pre- and post-processing operations \cite{shor1995,steane1996,terhal2015,Yu2007} or decoherence-free subspaces \cite{zanardi1997,lidar1998}, is clearly ineffective since entanglement cannot be created out of separable states exploiting only local operations and classical communication. Other techniques introduced in order to cope with decoherence and dissipation, such as dynamical decoupling and bang-bang control techniques \cite{viola1998,viola2005, damodarakurup2009}, while being potentially convenient for preserving static quantum memories \cite{biercuk2009}, are instead impractical for systems physically traveling along a quantum channel. Indeed such methods would require to continuously modulate the Hamiltonian of the system during the transmission line where usually Alice and Bob have no direct access, moreover these techniques are effective only in the non-Markovian regime in which the relaxation time of the environment is larger the time-scale of the control operations. Non-Markovian memory effects have also been used in \cite{orieux2015} to restore entanglement, but this is possible only if the environment degrees of freedom are directly accessible to Bob.

\begin{figure}[t]
\includegraphics[width=1.0 \columnwidth]{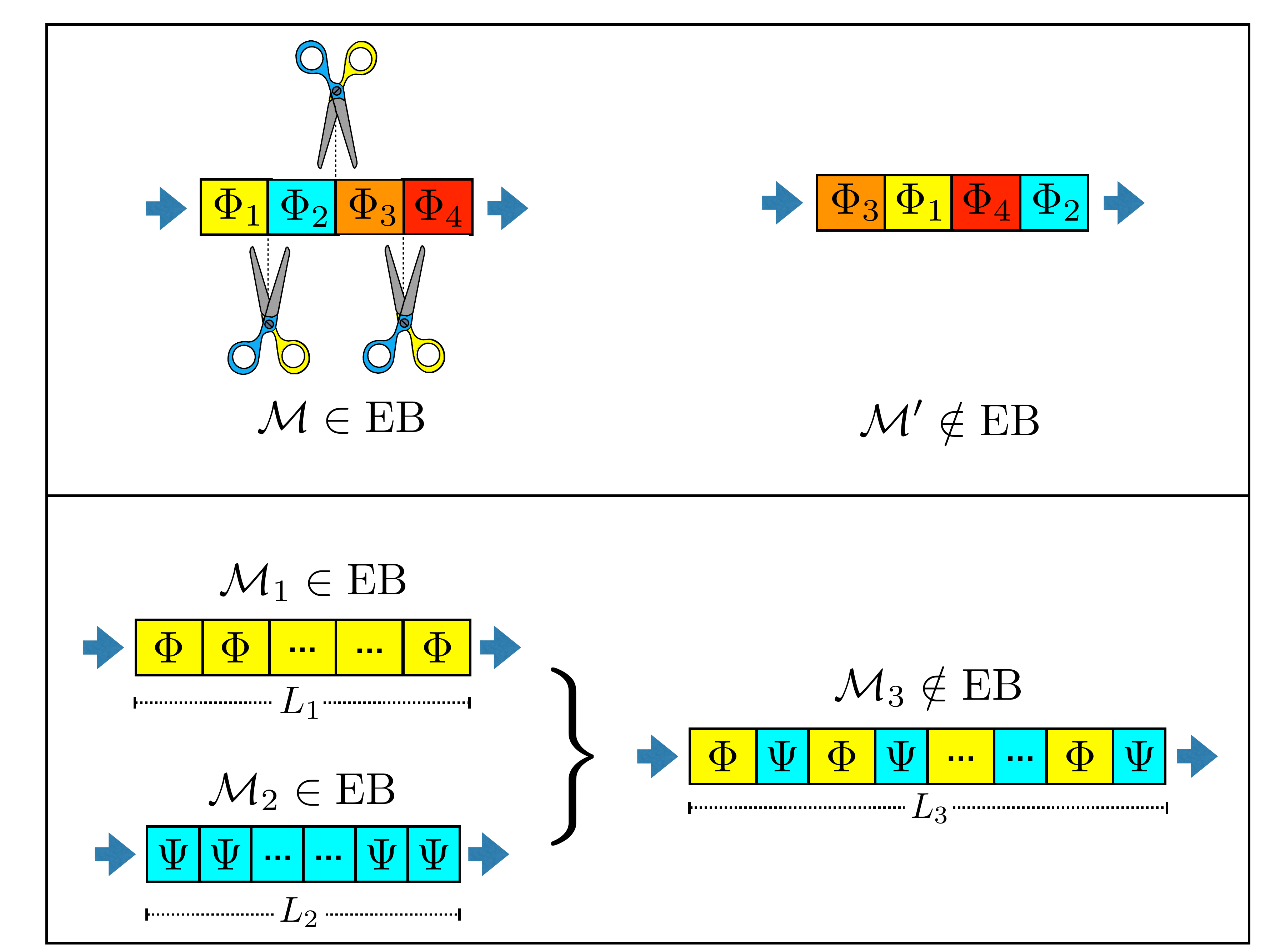}
\caption{Upper panel: Pictorial representation of the cut-and-paste technique. The original communication channel ${\cal M}$ is entanglement breaking. By dividing 
it in smaller sections represented by the maps $\Phi_1$, $\Phi_2$, $\Phi_3$, $\Phi_4$ and recombining them in a proper order we construct a new communication line ${\cal M}'$ 
which is less noisy than the original one, in particular it is not entanglement breaking. Lower panel: merging of two spatially homogeneous EB channels ${\cal M}_1$ and ${\cal M}_2$  to form the non EB map ${\cal M'}$.  } 
\label{fig:forbici}
\end{figure}

In this work we present a new approach, which we dub   {\it cut-and-paste},  that focuses on the way a given channel is assembled proving
that is possible to build reliable communication lines by  starting from extremely noisy components.
Specifically we show, both theoretically and experimentally, that it is possible to restore the transferring of quantum correlations along a communication line of fixed length, 
by simply splitting  it into smaller pieces and by re-organizing them  to form a  new channel of the same
length but which is less noisy than the original one.   
The idea is exemplified in  the upper panel of Fig.~\ref{fig:forbici}. Here  the map ${\cal M}$ 
associated with  a communication line that connects two distant parties   over a certain distance is
represented as an ordered sequence of smaller elements, i.e. ${\cal M} = \Phi_N \circ \Phi_{N-1} \circ \cdots \circ \Phi_2 \circ \Phi_1$, where for $n=1, \cdots, N$, $\Phi_n$ is the
transformation which propagates the messages on the $n$-th section of the line. Now assume that ${\cal M}$ is EB: accordingly it will prevent the transferring of any form of quantum correlations. Yet, in what follows we are going to show that there are cases where, by simply reshuffling the order in which the sub-channels are connected with each other, we can create a new physical map ${\cal M}' = \Phi_{i_N} \circ \Phi_{i_{N-1}} \circ \cdots \circ \Phi_{i_2} \circ \Phi_{i_1}$ that doesn't suffer from such limitations ($i_1,i_2, \cdots, i_N$ being a permutation of  the first $N$ natural numbers). An interesting  application of this effect is presented in the lower panel of Fig. \ref{fig:forbici}. Here we are in the presence of two communication lines, ${\mathcal M}_1$ and ${\mathcal M}_2$, 
of length $L_1$ and $L_2$ respectively, which  are both EB and which, for the sake of simplicity, we assume to be spatially homogenuous. As we shall see in the next sections, there are situations in which we can create a new communication line ${\mathcal M}_{3}$  which allows one to reliably propagate quantum coherence over distances $L_3$ much larger than $L_1 + L_2$, by simply alternating pieces of the original maps
i.e.
\begin{eqnarray} \label{MERGE} 
{\cal M}_{3}=( \Phi \circ \Psi )\circ ( \Phi \circ \Psi)  \circ \cdots \circ  (\Phi \circ \Psi) \notin \mbox{EB} 
\end{eqnarray} 
with $\Phi$ (resp. $\Psi$) being the sub-channel that composes ${\cal M}_1$ (resp. ${\cal M}_2$) -- the maximum value of $L_3$ being a function of the size of the pieces we have selected.

 The cut-and-paste effect detailed above ultimately relays on the non-commutative character of quantum mechanics, where it is not just the kind of operations performed on a system that matters, but also the order in which  they are carried on. At variance with previous applications, here such peculiar aspect of the theory exhibits its full potential impact,  by allowing or not-allowing  a well defined operational task (i.e. the sharing of entanglement among distant parties). 
On a more practical ground, our results can find applications in the realization of quantum networks connected by highly damping and noisy communication links. More generally
they suggest a new way of engineering quantum devices, widening the possibilities
 of optimising their performances by simply reordering the elements which constitute them.

In what follows we review some basic facts about EB maps and discuss the cut-and-paste mechanism by focusing on two examples: the first
regards amplitude damping and phase damping  channels operating on a qubit~\cite{nielsen2010}; the second instead deals with a continuous 
 model in which the propagation of signals in a noisy environment is represented in terms of effective master equations. 
Then for the amplitude damping example we present an experimental test of this effect where the qubits are encoded into the polarization degree of freedom of single photons.\\

\section{Theoretical analysis}

A quantum channel $\Phi$
is said to be EB  if, when applied to one part of an entangled state, the output is always separable for every choice of the input~\cite{horodecki2003}. Moreover it can be shown that   ${\Phi} \in \mathrm{EB}$   if and only if, when operating locally on the system of interest $S$, it turns a maximally entangled state $|\Omega\rangle_{SA}$ of $S$ and of an ancillary system $A$, into a separable one \cite{horodecki2003}, {\it i.e.}:
\begin{equation}
\rho_{SA}^{(\rm out)}=\left ( \Phi \otimes \mathbb{I} \right )(|\Omega \rangle_{SA}\langle \Omega |) \; \textrm{is separable}  \Longleftrightarrow   \Phi \in \mathrm{EB}\;, \label{test} 
\end{equation}
with $\mathbb{I}$ being the identity channel operating on $A$. A possible way to smooth the boundary between these ``bad" kind of channels and the 
``good" non-EB channels, is the notion of EB {\it order} of a channel introduced in \cite{depasquale2012}, and further studied in \cite{lami2015}. Accordingly  $\Phi$ is  said to be EB of order $n$ if it requires $n$ consecutive applications to destroy the entanglement of any input state, {\it i.e.} if up to $n-1$ iterations of $\Phi$ are not EB, while $n$ (or more) iterations of $\Phi$ yields a EB transformation. 
\\

\subsection{Cut-and-paste with discrete maps}
Amplitude Damping (AD) channels are an important  class of maps acting on a qubit~\cite{nielsen2010}. They   describe
dissipation and decoherence processes of a two-level system in thermal contact with a zero temperature external bath, or  the loss of a photon in the propagation of single photon pulses along an optical fiber.   Given an input density matrix $\rho$ the channel $\mathcal{A}_\eta$ transforms it as 
$\mathcal{A}_\eta(\rho)\equiv E_1 \rho E_1^\dagger +E_2 \rho E_2^\dagger$,
where
$E_1 \equiv
   \tiny{\left[ \begin{array}{cc}
  1& 0   \\
0 & \sqrt{\eta}  \end{array} \right]}$ and 
$E_2 \equiv \tiny{\left[ \begin{array}{cc}
  0&  \sqrt{1-\eta}   \\
 0  & 0  \end{array} \right]}$,
are the Kraus operators expressed in the computational basis of the qubit, and $\eta\in [0,1]$ is the transmission coefficient characterizing the map, interpolating between perfect transmission $\eta=1$ and complete damping $\eta=0$. One can easily verify that $\mathcal{A}_\eta$ is never EB for $\eta>0$ and,  from the semigroup property 
\begin{eqnarray} 
\mathcal A_{\eta_2} \circ \mathcal A_{\eta_1} = \mathcal A_{\eta_1} \circ \mathcal A_{\eta_2} = \mathcal A_{\eta_2\eta_1} \label{semi},
\end{eqnarray} 
that it also has  infinite order, i.e. $n=\infty$. Interestingly enough however, it can be shown \cite{depasquale2012} that for sufficiently small values of the transmission coefficient $\eta$, the order of $\mathcal A_\eta$ can be reduced by post-posing or ante-posing  suitable unitary gates. In particular one can identify  (non-unique) unitary operations $\rho \rightarrow \mathcal U(\rho) =U \rho U^\dag$ such that the channels
\begin{eqnarray}\label{UdagA} 
\Phi=\mathcal A_\eta \circ \mathcal U , 
 \qquad 
\Psi=\mathcal U^\dag \circ A_\eta,
\end{eqnarray}
are both EB of order $n=m=2$: 
 for instance this happens at $\eta =0.3$ when we take the bit-flip $\sigma_x$ as $U$. 
 What it is more interesting, we can now use these maps to provide a first evidence of  the cut-and-paste mechanism. 
 Indeed consider the compound  map ${\cal M}$ formed by two applications of $\Phi$ and followed by two applications of $\Psi$, i.e. 
 ${\cal M} = \Phi\circ \Phi \circ \Psi \circ \Psi$ (corresponding to set  $\Phi_1=\Phi_2=\Phi$ and $\Phi_3=\Phi_4= \Psi$ in the upper panel of  Fig.~\ref{fig:forbici}). This is clearly EB, however 
  the alternate application of the two maps produces the AD channel of transmissivity $\eta^4$  which, as already stated,  is never EB:
\begin{eqnarray}\label{perfect_cut&paste}
&&{\cal M}' = \Phi \circ \Psi \circ \Phi \circ \Psi \\ &&\quad = 
 \mathcal A_\eta \circ \mathcal U   \circ \mathcal U^\dag \circ  \mathcal  A_\eta  \circ  \mathcal A_\eta \circ \mathcal U   \circ \mathcal U^\dag \circ \mathcal A_\eta   \nonumber =  \mathcal A_{\eta^4} \notin \mathrm{EB}.
\end{eqnarray}
Similar results can also be obtained by 
replacing in the above expressions  the AD map $ \mathcal A_\eta$  with the phase-damping (PD) channel $\mathcal{E}_p$~\cite{nielsen2010}. For $p\in [0,1]$
the latter  transforms a generic input density matrix $\rho$  of the qubit as $\mathcal{E}_p(\rho)\equiv 1/2 \left((1+p) \rho   + (1-p)  \sigma_z \rho \sigma_z \right)$ describing  
loss of coherence  between its energy eigenstates.
Analogously to the AD maps,  $\mathcal{E}_p \notin \mathrm{EB}$ for $p>0$ and satisfies the semigroup property $\mathcal E_{p_2} \circ \mathcal E_{p_1} = \mathcal E_{p_1} \circ \mathcal E_{p_2} = \mathcal E_{p_2 p_1}$, from which it immediately follows that they are EB of  infinite order, i.e. $\mathcal{E}_p \in \mathrm{EB}^n$ for $n=\infty$. Furthermore also in this case for $p$ sufficiently small one can find a unitary transformation $\cal U$ such that the maps $\Phi=\mathcal{E}_p \circ \mathcal U$ and $\Psi=\mathcal U^\dagger \circ \mathcal{E}_p $ are  EB of order 2, e.g. by fixing $p=0.4$ and $U=1/\sqrt{2}(\sigma_z - \sigma_x)$.

 The examples presented  here constitute also  a clear ``proof of principle" demonstration of the merging effect described in the lower panel of Fig.~\ref{fig:forbici}. 
   Indeed both in the AD and in the PD implementation, 
not only the channel ${\cal M}$ is EB, but also the maps associated with its first and second sections (i.e. the maps  ${\cal M}_1= \Phi\circ\Phi$   and ${\cal M}_2=\Psi\circ \Psi$ respectively) 
share the same property. Yet even though individually such transformations 
are both bad operations, when split and merged as in~(\ref{perfect_cut&paste}) they manage to somehow correct
their detrimental effects. 
This is a curious instance of an effective quantum error correction procedure obtained by properly mixing two different types of errors. 
Furthermore, due to their semigroup property which both the AD and PD maps fulfil, 
the channel $\Phi \circ \Psi$
has actually an infinite order ($n=\infty$). Accordingly the procedure can be 
iterated an arbitrary number of times obtaining an infinitely long sequence $(\Phi \circ \Psi )\circ (\Phi \circ \Psi )\circ  \dots$ which nonetheless can still transmit a non-zero fraction of entanglement (corresponding to have a divergent value of $L_3$ of Fig.~\ref{fig:forbici}). \\

\subsection{Cut-and-paste with continuous channels}
The potentiality of the cut-and-paste method is much more general and 
is not restricted to idealized sequences of discrete unitary and dissipative operations discussed so far. In more realistic scenarios, like the continuous propagation of a quantum state along a physical medium, the transmission is better 
described by continuos homogeneous channels (also known as quantum dynamical semigroups) \cite{QDsemi, masterEq}. In this framework, the propagation of a signal for an infinitesimal distance $dx$, induces a change of the density matrix 
according to a master equation of the form:
\begin{equation}\label{master1}
\frac{d \rho}{dx}= \mathcal L ( \rho),
\end{equation}
where the linear Liouvillian operator $\mathcal L$, completely characterizes the channel and generates simultaneously the unitary and dissipative dynamics of the system \cite{QDsemi, masterEq}. 
The formal solution of Eq. \eqref{master1}, is $\rho(x)= e^{\mathcal L x} \rho(0)$ and represents the propagation of the quantum state for a finite length $x$ along the physical medium.

Now, as in the case of the lower panel of Fig.~\ref{fig:forbici}, assume that we have at disposal two of these maps ({\it e.g.} two different wave guides) characterized  by $\mathcal L_1$ and $\mathcal L_2$, such that 
the integrated dynamics becomes EB after a propagation length of $x=L_1$ and $x=L_2$ respectively, {\it i.e.} :
\begin{equation}\label{master3}
{\cal M}_1= e^{\mathcal L_1 L_1} \in {\rm EB},    \qquad  {\cal M}_2=e^{\mathcal L_2 L_2} \in {\rm EB} .
\end{equation}
If we divide the two transmission lines into shorter pieces of length $L_1/n_1$ and $L_2/n_2$ we, obtain the following maps
 $\Phi = e^{\mathcal L_1 L_1 / n_1}$ and 
  $\Psi  =    e^{\mathcal L_2 L_2 / n_2}$, 
which are, by construction, EB of order $n_1$ and $n_2$.  The idea is thus to construct a new communication line 
using $\Phi$ and $\Psi$ as elementary building blocks which are merged to form an alternate sequence  as in Eq.~(\ref{MERGE})  with the aim of
increasing the entanglement propagation length as much as possible. 
As an example we can consider two hypothetical transmission lines, characterized by the following Liouvillian operators \cite{QDsemi, masterEq}:
\begin{equation}\label{masterAD}
\mathcal L_j(\rho) = -{i} [H_j, \rho] + \frac{\epsilon}{2} ( 2 \sigma_{-} \rho \sigma_{+} -  \sigma_{+}  \sigma_{-} \rho -\rho \sigma_{+}  \sigma_{-} ),
\end{equation}
where $j=1,2$ ,  $H_1= \Omega \sigma_x$, $H_2= - \Omega  \sigma_x$ and $\sigma_{\pm} = (\sigma_x \pm i \sigma_y)/2$. 
The first term in this equation 
induces a unitary rotation of the quantum state while the second term is responsible for the dissipative dynamics. 
Accordingly  Eq.~(\ref{masterAD}) is  a sort of continuous analogue of the  discrete maps of Eq.~(\ref{UdagA}) where the two different form of transformations (i.e. $\mathcal A_\eta$ and $ \mathcal U$) are 
 applied to the system not one after the other but simultaneously. 

For nonzero values of $\Omega$ and $\epsilon$, both channels become EB at the same finite propagation length $L$. 
For simplicity we cut slices of equal length $L/n$ for both channels, and we paste them according to the alternating sequence given in Eq. \eqref{MERGE}.
This new continuous channel is formally described by a master equation in which the Liouvillian operator switches between $\mathcal L_1$ and $\mathcal L_2$ after the propagation of $L/n$ space intervals, {\it i.e.}

\begin{equation} \label{master12}
\frac{d \rho}{dx}= \mathcal L_3^{(n)} (\rho)= {\Bigg \{ }
\begin{array}{ll}
\mathcal L_1(\rho), \quad {\rm for}\;    \big[ \frac{x}{L/n} \big]= {\rm  even}, \\
\\
\mathcal L_2(\rho), \quad {\rm for}\;   \big[ \frac{x}{L/n} \big]= {\rm  odd},
\end{array}
\end{equation}
where $[\cdot]$ is the integer part of the argument.

\begin{figure}[t]
{\includegraphics[width=0.98 \columnwidth]{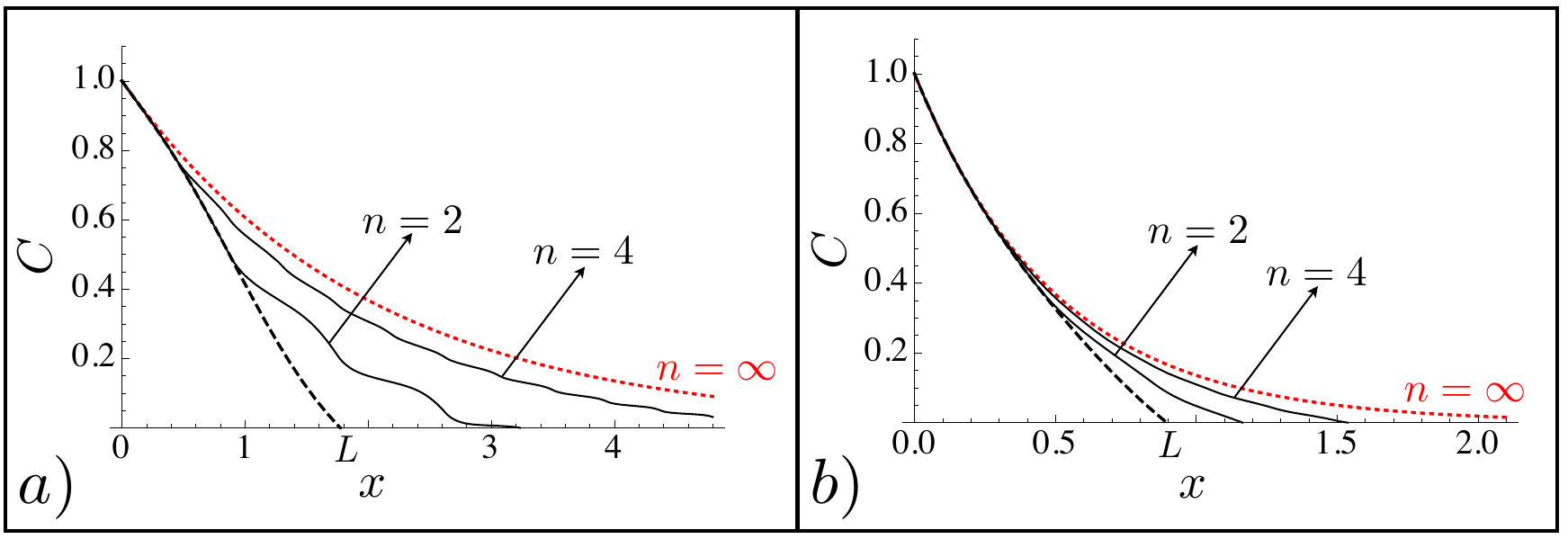}}
\caption{Residual concurrence~\cite{CONC} as a function of the propagation distance $x$ for different continuous channels. At $x=0$ the system is initialized in the singlet state. Panel a) refers to the rotating amplitude damping channels defined in Eq.~\eqref{masterAD} (black dashed line) and their associated cut-and-paste improvements described in Eq.~\eqref{master12} for different values of $n$ (black lines). The limit for $n\rightarrow \infty$ is also represented (red dotted line). Panel b) instead is based on the rotating phase damping channels defined in Eq. \eqref{masterPD}, while the notation is the same as for panel a). The values of the parameters are $\Omega=1.5$ and $\epsilon=1$ for both subfigures which yield $L=1.75$ for Eq.~(\ref{masterAD}) and $L=0.85$ for Eq.~(\ref{masterPD}) for the maximum propagation length that allow for quantum coherence preservation (beyond these values the maps become EB). The cut-and-paste approach allows to significantly increase the EB threshold well beyond the original one: for instance in the case of Eq.~(\ref{masterAD}) setting $n=2$ already gives $L_3$ which is more than twice the value of $L$ (similarly this happens also for the case of  Eq.~(\ref{masterPD}) by setting $n=4$).  }
\label{fig:master}
\end{figure}

We now test the entanglement condition given in Eq.\ \eqref{test}, {i.e.}  we study the entanglement evolution induced by the channel $\mathcal L_3^{(n)}$ when applied to one part of a maximally entangled state.
The value of entanglement measured in terms of the concurrence~\cite{CONC} as a function of the distance $x$ is represented in Fig. \ref{fig:master}a  for different values of $n$. We observe that with respect to the original channels $\mathcal L_1$
 and $\mathcal L_2$, the new channels $\mathcal L_3^{(n)}$ become entanglement breaking at larger distances. Moreover, the entanglement propagation length increases with $n$, and tends to infinity. This fact can be 
proved theoretically: using a simple Trotter decomposition argument, the dynamics for $n \rightarrow \infty$ tends to a pure amplitude damping channel which is never entanglement breaking.

Exactly the same analysis and similar results are valid also for other continuous channels. For example if we replace equation \eqref{masterAD} with
\begin{equation}\label{masterPD}
\mathcal L_j(\rho) = -{i}[H_j, \rho] + \frac{\epsilon}{2} [\sigma_z ,  [\sigma_z,\rho]  ],
\end{equation}
we obtain two propagation media indexed by $j=1,2$ in which a qubit is rotated in different directions by $H_1= \Omega \sigma_x$ and $H_2= - \Omega  \sigma_x$  while, at the same time, it is subject to a dephasing process \cite{QDsemi,masterEq}.
Remarkably, this is a quite realistic model for the propagation polarization qubits in optical fibers, in which dephasing and polarization drift can simultaneously affect the transmitted photons. Fig. \ref{fig:master}b demonstrates that our approach is applicable also in this situation, obtaining a significant enhancement of the entanglement propagation distance.\\

\begin{figure*}[t!]
\includegraphics[width=1.5 \columnwidth]{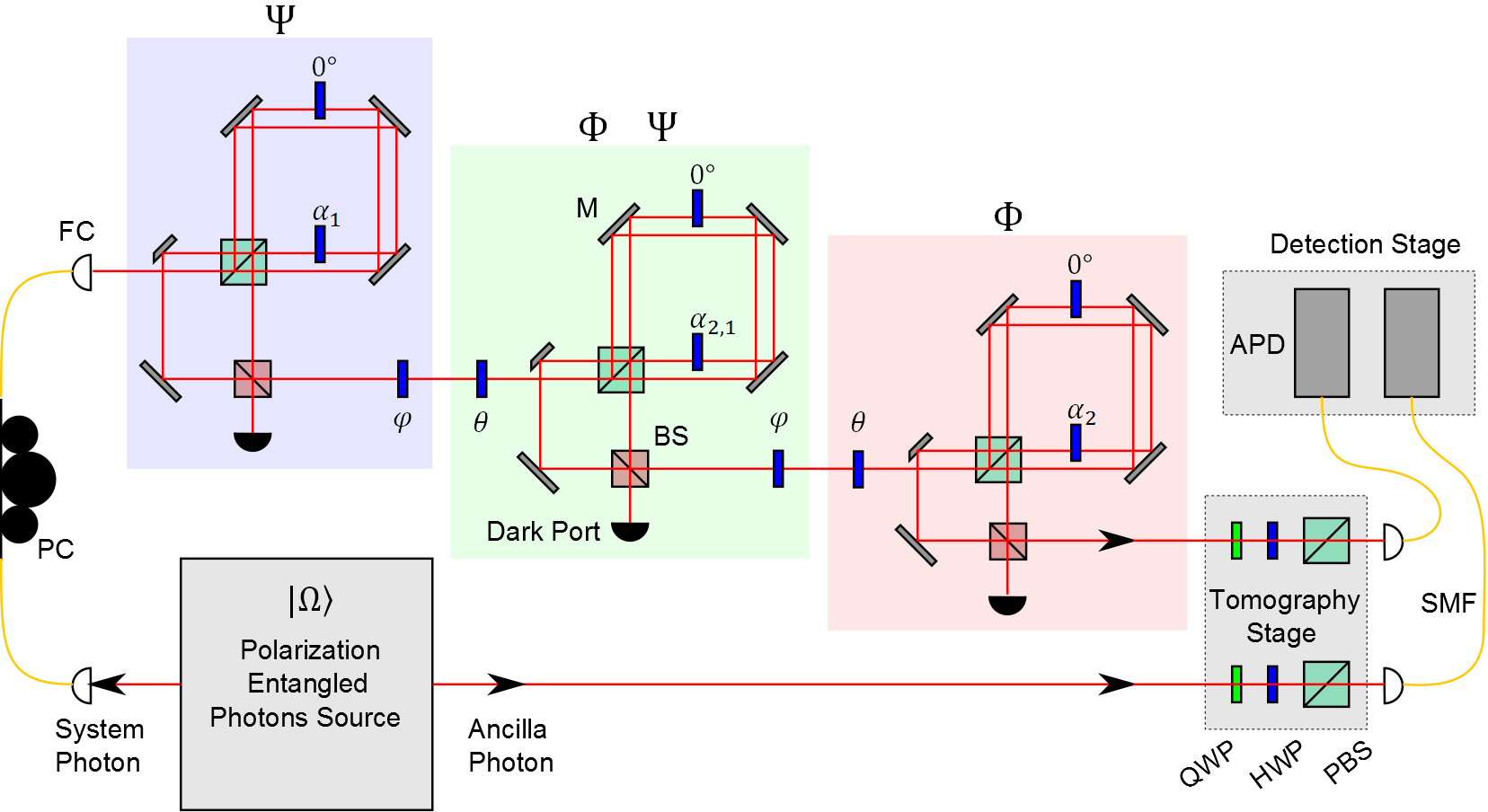}
\caption{Scheme of the experimental setup. The $A$ photon belonging to the polarization entangled state $|\Omega\rangle_{SA}$ is directly linked to the detector, while the $S$ photon is connected to the three DIF via a single-mode optical fiber (SMF). A bipartite full tomography stage, given by one HWP and one quarter wave plate (QWP) for both the system and the ancilla photons, allows to reconstruct the density matrix of the two-photon state. Finally, photons pairs are coupled to two electronically synchronized avalanche photodiodes (APDs) through two SMFs. PC: Polarization Controller. BS and PBS: Beam Splitter and Polarizing Beam Splitter.}
 \label{fig:Scheme}
\end{figure*}

\section{Experimental implementation}

In this section we present a quantum optics experiment demonstrating the possibility of restoring the transmission of entanglement via the cut-and-paste technique previously described. Specifically we study the transmission of a qubit encoded in the horizontal and vertical polarization of a photon $\{ |0 \rangle\equiv | H \rangle, |1 \rangle\equiv |V\rangle \}$ 
and, inspired by Eq.~\eqref{UdagA}, we take
\begin{eqnarray} 
\Phi =  \mathcal A_{\eta_2}  \circ {\cal U}_\theta \;, 
 \qquad 
\Psi =  {\cal U}_\varphi  \circ   \mathcal A_{\eta_1}\;,
 \label{psi2expAmp}
\end{eqnarray}
where now the unitary mappings ${\cal U}_{\xi}$ with $\xi=\theta,\varphi$ induce  the rotation
$U_{\xi} \equiv \tiny{\left[ \begin{array}{cc}
  \cos(2\xi)&  \sin(2\xi)   \\
  \sin(2\xi)  & -\cos(2\xi)  \end{array} \right]}$ in the polarization degree of freedom of the photon.   
	
The experimental setting is sketched in Fig.~\ref{fig:Scheme}: for assigned values of $\eta_1$, $\eta_2$, $\theta$ and $\varphi$, it is designed to study the presence/absence of entanglement on a maximally polarization entangled state $|\Omega\rangle_{SA}$ of a two-photon pair, which evolves under the transformation~(\ref{test}) given by the channels ${\cal M}'=\Phi \circ \Psi \circ \Phi \circ \Psi$, ${\cal M}_1=\Phi\circ \Phi$, or ${\cal M}_2=\Psi\circ \Psi$ by properly tuning the parameters of the interferometer (see below for details). 
For each one of these three choices the EB character of the transformation  is hence determined exploiting the equivalence~(\ref{test}) by measuring the concurrence~\cite{CONC} of the associated density matrix  $\rho_{SA}^{(\mathrm{out})}$  through full tomography performed at the output of the setup: evidence of the cut-and-paste effect is thus obtained whenever one notices a non-zero value for the concurrence for
${\cal M'}$ in correspondence with a zero concurrence value for  ${\cal M}_{1}$ and ${\cal M}_{2}$.
 
In our test we employ as $|\Omega\rangle_{SA}$ the superpositions $(|H\rangle_{S}|V\rangle_{A}+ e^{i\phi}|V\rangle_{S}|H\rangle_{A})/\sqrt{2}$ created through a high-brilliance, high-purity polarization entanglement source (see \cite{Fedrizzi}). The source consists of a nonlinear PPKTP crystal pumped by a single mode laser at $405nm$ and $2.75mW$ of power within a Sagnac Interferometer (SI), and able to generate pairs of photons in the system mode  $S$ and the ancillary mode $A$, at $810nm$ by Type-II parametric down conversion. The generated pairs (more than 50000 detected coincidences/sec) have a coherence length of $L_{coh}=1.02mm$ and spectral bandwidth $\Delta \lambda=0.43nm$. The ancillary photon $A$ is hence directly transmitted to a detector station, while the photon $S$ is connected to the testing area which implements the action of  the  maps ${\cal M}'$, ${\cal M}_1$ and ${\cal M}_2$ through both free-space and fiber optics links, and then sent to a second detector station, see Fig.~\ref{fig:Scheme}. 

When emerging from the source, the resulting state has more than $98\%$ of fidelity with the target $|\Omega\rangle_{SA}$ and concurrence equal to $C=0.98$, corresponding to an effective $C=0.94$ at the interferometer detectors, because of the unavoidable entanglement degradation occurring in the connection with the testing channel (see Appendix~\ref{appendixB}).
  
The mappings ${\cal U}_{\xi}$ entering~(\ref{psi2expAmp}) are implemented by placing properly oriented Half-Wave Plates (HWP) along the optical axis of  the propagating photons.  The AD channels ${\cal A}_{\eta_1}$ and ${\cal A}_{\eta_2}$ are instead realized by means of a dual interferometric setup (DIF) obtained by putting two independent polarization controls inside a displaced SI, coupled to an external unbalanced Mach-Zehnder interferometer (MZI) (see Appendix~\ref{appendixA} for more details on this). 

To implement ${\cal M'}=\Phi \circ \Psi \circ \Phi \circ \Psi$ we exploit the semigroup property~(\ref{semi}) to formally express the product $\mathcal A_{\eta_{1}} \circ \mathcal A_{\eta_{2}}$ as a single AD channel with damping parameter $\eta_{2}\eta_{1}$. Accordingly we write 
\begin{eqnarray} 
{\cal M}'=  \mathcal A_{\eta_2}  \circ {\cal U}_\theta  \circ  {\cal U}_\varphi  \circ   \mathcal A_{\eta_1\eta_2}
  \circ {\cal U}_\theta \circ  {\cal U}_\varphi  \circ   \mathcal A_{\eta_1}. \label{sequence} 
\end{eqnarray} 
This simple trick enables us to implement the whole transformation by means of three DIFs only instead of four (one for each AD map entering the original sequence), by taking  the parameters $\alpha_1$, $\alpha_{2,1}$ and $\alpha_2$ of Fig.~\ref{fig:Scheme} equal to $\arccos(-\sqrt{\eta_1})/2$, $\arccos(-\sqrt{\eta_2\eta_1})/2$, and $\arccos(-\sqrt{\eta_2})/2$, respectively (see Eq.~(\ref{alphadef}) of the Appendix~\ref{appendixA}). For fixed values of $\eta_2$ and $\theta$, the action of ${\cal M}_{1} =\Phi\circ \Phi$ is studied by deactivating both $\Psi$ maps in~(\ref{sequence}). Such operation is realized in two steps. First, the HWP placed in the internal vertically polarized path of the first DIF of Fig.~\ref{fig:Scheme} is set to $\alpha_1=\pi/2$ to simulate $\eta_1=1$, while the HWPs placed in the internal vertically polarized paths of the second and third DIF, are set to induce the same rotation (i.e. $\alpha_{2,1}=\alpha_2=\arccos(-\sqrt{\eta_2})/2$).  Secondly, both the external HWPs of Fig.~\ref{fig:Scheme} which are responsible for the implementation of $\mathcal U_{\varphi}$ are simply taken off from the setup. Similarly, for each given value of $\eta_1$ and $\varphi$, we can study the action of ${\cal M}_{2}=\Psi\circ \Psi$ deactivating both $\Phi$ maps, by taking off from the setup both the external HWPs associated with the rotation $\mathcal U_{\theta}$, and by letting the internal HWPs of the three DIFs as $\alpha_1=\alpha_{2,1}= \arccos(-\sqrt{\eta_1})/2$ and $\alpha_2=\pi/2$ to simulate $\eta_2=1$. Finally, the implementation of the identity channel (useful to directly measure the net entanglement of the input state available at the detectors of the interferometer) is obtained by setting $\alpha_1=\alpha_{2,1}=\alpha_2=\pi/2$ and removing all the external HWPs.\\

 \begin{figure}[t]
\includegraphics[width=0.98 \columnwidth]{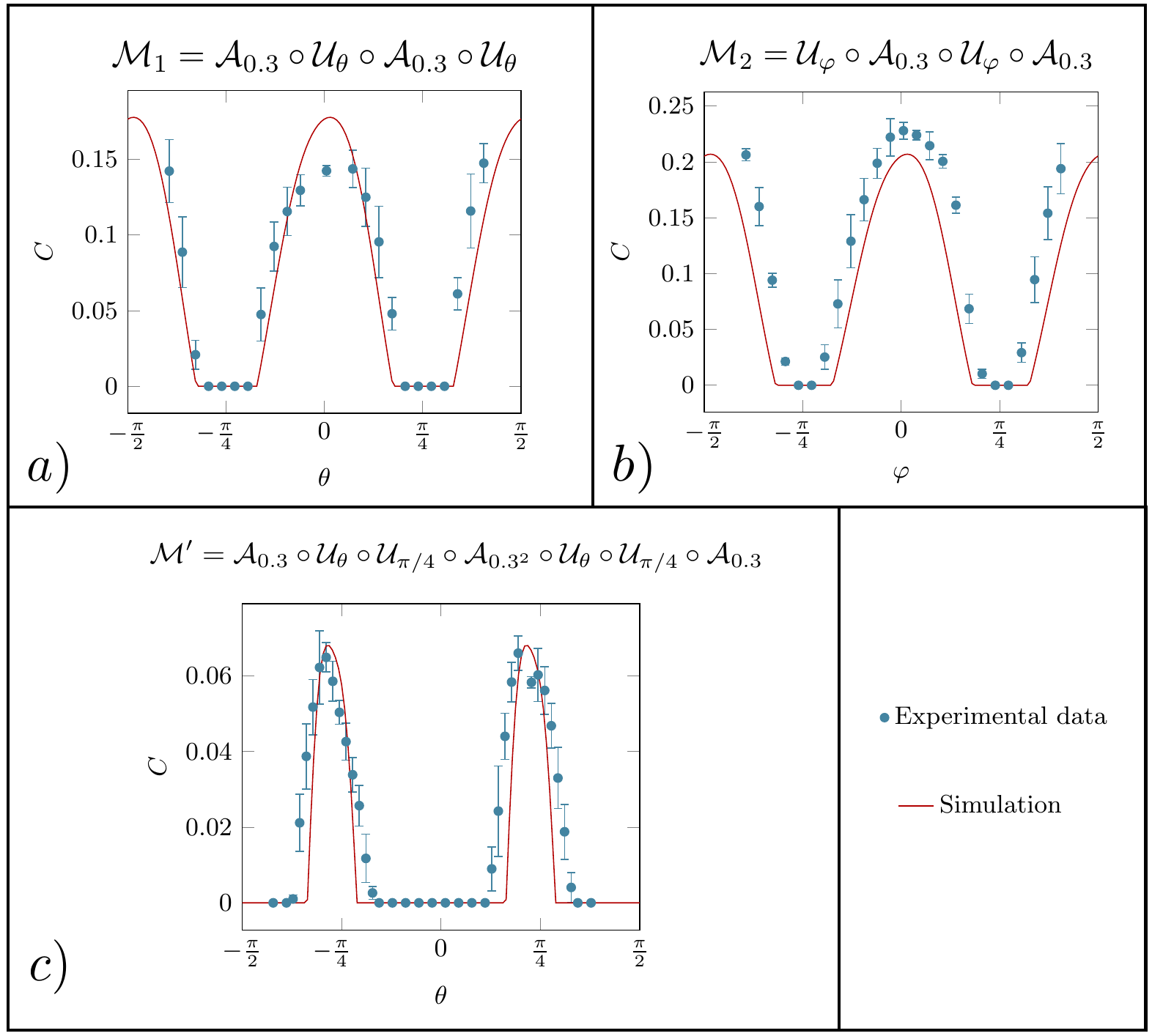}
\caption{Entanglement breaking analysis for the maps ${\cal M}_{1}=\Phi\circ\Phi$, ${\cal M}_{2}=\Psi\circ\Psi$ and ${\cal M'}=\Phi\circ\Psi\circ\Phi\circ\Psi$ 
 obtained by cascading the channels of Eq.~(\ref{psi2expAmp}) with $\eta_1=\eta_2=0.3$. 
a)  Functional dependence upon the angle $\theta$ of the concurrence ${C}$  of the  state (\ref{test}) at the output the channel ${\cal M}_1$  obtained from the scheme of Fig.~\ref{fig:Scheme} by setting $\alpha_1=\pi/2$, $\alpha_{2,1}=\alpha_2=\arccos(-\sqrt{\eta_2})/2$ and removing the rotations $\mathcal U_{\varphi}$ (see text). For  $\theta$ around $\pm\pi/4$ the system exhibits zero concurrence showing that the $\Phi\circ \Phi$ is EB, or equivalently that $\Phi$ is EB of order 2. 
b) Functional dependence upon the angle $\varphi$ of the concurrence ${C}$  of the  state (\ref{test}) at the output the channel ${\cal M}_{2}$ obtained  by  setting $\alpha_1=\alpha_{2,1}= \arccos(-\sqrt{\eta_1})/2$ and $\alpha_2=\pi/2$ and removing the rotations $\mathcal U_{\theta}$.
Similarly to the  previous case, for $\varphi$ around $\pm\pi/4$ the system exhibits zero concurrence showing that the $\Psi\circ \Psi$ is EB, or equivalently that $\Psi$ is EB of order 2. 
c)  Functional dependence upon the angle $\theta$ of the concurrence ${C}$  of the  state (\ref{test}) at the output the channel ${\cal M}'$ obtained
by setting $\alpha_1=\alpha_2=\arccos(-\sqrt{0.3})/2$, $\alpha_{2,1}=\arccos(-0.3)/2$ and 
keeping $\varphi=\frac{\pi}{4}$. In all plots the red curves represents the theoretical prediction obtained by considering the actual experimental conditions (see Appendix~\ref{appendixB}). Each point and the associated statistical error was taken from a set of $N$ measurements ($3 \le N \le 11$), under equivalent mode coupling conditions.}
 \label{fig:Cut_and_pasteNEW}
\end{figure}

In Fig.~\ref{fig:Cut_and_pasteNEW} we report the results obtained having set the transmission coefficients of the AD channels entering Eq.~(\ref{psi2expAmp}) at $\eta_1=\eta_2 = 0.3$ (value for which we have already anticipated  the possibility of witnessing the cut-and-paste effect).
In particular panels $a)$ and $b)$ show the EB analysis for the maps ${\cal M}_{1}=\Phi\circ \Phi$ and ${\cal M}_2 = \Psi\circ\Psi$ when, respectively, $\theta$ and $\varphi$ are varied.
In both cases one may notice that the concurrences of the associated $SA$ output states~(\ref{test})  nullify when these angles reach the values~$\pm \pi/4$, implying hence that under these conditions the channels $\Phi$ and $\Psi$ are EB of order 2. 
Panel $c)$ shows instead  the EB analysis for the map  ${\cal M}'$ of Eq.~(\ref{sequence}) when the angle $\varphi$ is kept constant at $\frac{\pi}{4}$ and $\theta$ is varied: one notices that the concurrence of the output state~(\ref{test}) is peaked and different from zero around $\theta=\pm \pi/4$, indicating that for these values the map is not EB.
Our data  provide hence a clear experimental evidence of the cut-and-paste entanglement restoring effect for $\eta_1=\eta_2=0.3$,  $\theta=\pm\pi/4$, and $\varphi=\pm\pi/4$. 
The theoretical prediction corresponding to the red curves shown in the figure have been obtained by taking into account the actual optical elements of the experimental setup --  the slight residual disagreement being mainly due to the unavoidable difficulties of coupling different polarization and path contributions within the same single mode fiber (see Appendix~\ref{appendixB} for details). \\

\section{Conclusions}

``Cutting'' two entanglement-breaking channels {into two or more pieces} and properly reordering the corresponding parts can yield a new communication line which is not entanglement-breaking: this is the essence of the cut-and-paste protocol.  
In this work we give the first theoretical proposal {both for discrete and continuous time evolution, the latter proving a more realistic model for signals evolution within a piece of material. We considered as benchmarks for a proof-of-principle demonstration the rotated amplitude and phase damping maps. In the second part of the manuscript, we focused on the discrete amplitude damping evolution, yielding} the first experimental demonstration of this quite unconventional protocol. 
The quantum optics experiment that we have realized is based on the transmission of photon polarization qubits and has unambiguously proved the predicted entanglement recovery effect. The specific proof-of-principle demonstration has been constructed using combinations of amplitude-damping channels and unitary operations. However the proposed cut-and-paste technique is not limited to this scenario and can be applied also to more general single-qubit channels.  We also envisage interesting generalizations of our approach to arbitrary multi-qubit channels \cite{holevo2012} and to the physically important class of Gaussian entanglement-breaking channels \cite{holevo2008, depasquale2013}. 
Our technique and its experimental realization demonstrates the real possibility of recover some amount of entanglement or extend its distance distribution over extremely noisy links, opening a novel opportunity towards the realization of quantum information networks of increasing complexity \cite{duan2010,ritter2012}.
Moreover, the unusual entanglement recovery effect studied in this work is also interesting in its own right and could open new research lines in quantum channel theory \cite{holevo2012} and error correction protocols \cite{terhal2015}.

\section{Acknowledgements}
This work was supported by the ERC-Starting Grant 3D-Quest (3D-Quantum Integrated Optical Simulation; grant agreement no. 307783): http://www.3dquest.eu. It was also partially supported by the ERC through the Advanced Grant n. 321122 SouLMan, and by PhD Chilean Scholarships CONICYT, "Becas Chile".\\
 
\appendix
\section{Amplitude Damping Channel}\label{appendixA}

\begin{figure}[t]
\includegraphics[width=0.98 \columnwidth]{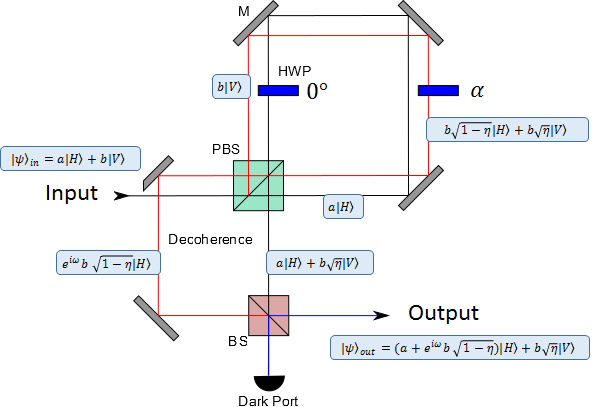}
\caption{Scheme of the experimental implementation of the AD map $\mathcal {A}_{\eta}$ via a DIF loop. In this picture the path {\bf a} associated with the $|H\rangle$ polarization component is indicated in black, while
the path {\bf b}  associated with the $|V\rangle$ component is in red. The blue elements of the figure represent HPW elements.  }
 \label{fig:amplitude_damping}
\end{figure}

In the polarization basis $\{|H\rangle,|V\rangle\}$ the action of the AD map $\mathcal {A}_{\eta}$ can be realized by means of the DIF of Fig.~\ref{fig:amplitude_damping}. Here an incoming signal $|\psi\rangle = a |H\rangle + b |V\rangle$ enters first a Sagnac Interferometer realized by means of a Polarizing Beam Splitter (PBS) which allows us to split the two polarization components by mapping them 
into two distinct optical paths, i.e.  the black path {\bf a} of Fig.~\ref{fig:amplitude_damping} for the $|H\rangle$ component and the red path {\bf b} for the $|V\rangle$ component, respectively.
Accordingly the state of the signal immediately after the PBS can be expressed as $a |H\rangle\otimes |\mbox{\bf a} \rangle + b |V \rangle\otimes | \mbox{\bf b} \rangle$ where we  expanded the Hilbert space  by explicitly adding the path degree of freedoms. Inserting then a rotated HWP along {\bf b}  a  rotation on the polarization degree of freedom of  $|V \rangle\otimes |\mbox{\bf b} \rangle$   can then be induced while leaving the  polarization of the $|H\rangle\otimes | \mbox{\bf a} \rangle$ unchanged, i.e.  $|V \rangle\otimes |\mbox{\bf b} \rangle \rightarrow \sqrt{\eta}|V\rangle\otimes |  \mbox{\bf b} \rangle+\sqrt{1-\eta} |H \rangle\otimes | \mbox{\bf b} \rangle$ and $|H\rangle\otimes | \mbox{\bf a} \rangle \rightarrow |H \rangle\otimes | \mbox{\bf a} \rangle$, the unitary operator
responsible for such transformation being 
\begin{eqnarray} \label{DDD}
U_{0}\otimes|\mbox{\bf a}\rangle\langle \mbox{\bf a}|+U_{\alpha(\eta)} \otimes|\mbox{\bf b}\rangle\langle \mbox{\bf b}|\;, 
\end{eqnarray}
where 
  \begin{eqnarray} 
\alpha(\eta) ={\arccos(-\sqrt{\eta})}/{2}\;, \label{alphadef}  
  \end{eqnarray} and where for $\alpha$ generic $U_\alpha$ indicates the polarization rotation associated with a HPW element rotated by $\alpha$, which in the basis $|H\rangle$, $|V\rangle$ is represented by the matrix 
\begin{equation}
U_\alpha=\left[ \begin{matrix}
  \cos(2 \alpha)&  \sin(2\alpha)   \\
  \sin(2 \alpha)  & -\cos(2\alpha)  \end{matrix}\right]\;, 
\end{equation}	
(as indicated in the Fig. \ref{fig:amplitude_damping} an unrotated HWP, performing the unitary transformation $U_0$ in Eq.~\eqref{DDD}, was also inserted in the  path ${\bf a}$ in order to preserve the temporal coherence between the two counter-propagating beams).
Subsequently, on their second encounter with the PBS, the two signals enter an unbalanced Mach-Zehnder interferometer which separates the horizontal component of  the {\bf b} path   adding to it a random phase $e^{i\omega}$ before recombining the signals at a $50\%$ Beam Splitter (BS). Accordingly, with probability $1/2$  the state emerging from the output port of the figure is described by the vector $|\psi\rangle_{out}=(a+e^{i\omega}b \sqrt{1-\eta})|H\rangle+b \sqrt{\eta}|V\rangle$ which, upon averaging over the random term $\omega$, corresponds to the action of $\mathcal {A}_{\eta}$  on $|\psi\rangle$, no photon emerging otherwise (in writing  $|\psi\rangle_{out}$ the path degree of freedom have been removed since the BS is effectively filtering out one of them).

In the experimental setting shown in Fig.~\ref{fig:Scheme} the above transformation is iterated three times in order to reproduce the sequences of the maps $\Psi$ and $\Phi$ as detailed in the main text, with the precaution of setting the physical dimensions of the associated  Mach-Zehnder interferometers to be different from each other in order to avoid any spurious coherence among the random phases they are meant to introduce.\\

\section{Simulation of the scheme}\label{appendixB}

The complexity of the geometry and the difficulties of coupling many spatial modes within one final single-mode fiber affect the entanglement preservation even if the maps implemented by the various DIF were set to operate as identity channels (i.e. taking $\eta_1=\eta_2 =1$ in Eq.~(\ref{sequence}) which implies setting $\alpha_1=\alpha_{2,1}=\alpha_2=0$, and physically removing the HWPs implementing the unitary rotations ${\cal U}_{\varphi}$ and ${\cal U}_{\theta}$). Indeed, under such experimental conditions, the mean entanglement degradation on each DIF was greater than $1.3\%$, thus the maximum concurrence at end of the entire sequence of channels decreases from $98\%$ to $94\%$. Besides the decrease of entanglement, the number of coincidences/sec was also affected, considering that the effective photon transmission of each double interferometer was $\approx1/3$.  Furthermore the effective operation of each interferometer, and therefore of each channel, can differ considerably from the others, even for small differences of the optical elements. 

The numerical simulations presented in Fig.~\ref{fig:Cut_and_pasteNEW}
  have been obtained by taking into account all these imperfections. 
In particular we  considers as input of the setup
 an effective input state of Werner form $\rho_{SA}=W |\Omega\rangle_{SA} \langle\Omega| +(1-W)\mathbb{I}_{SA}/4$ with the parameter $W\in[ 0,1]$ extrapolated from the degree of entanglement of the source. 
Regarding the BS transformations  instead we described  them as the following $2\times 2$ matrices with respect to the spatial degree of freedom associated with the input ports,
\begin{equation}
BS=\left[\begin{matrix}
\sqrt{T'} & i\sqrt{R'}\\
i\sqrt{R'} & \sqrt{T'}
\end{matrix}\right]\;, \label{MATR}
\end{equation}
where the measured optical transmissivity $T$ and reflectivity $R$ have been renormalized to include  losses $L = 1 -T-R$ by setting 
$T'=T/(1-L)$ and $R'=R/(1-L)$ (the average values of our set of BSs being $T=0.48$,$R=0.44$ and $L=0.08$).
Analogously to simulate the PBSs, we separated the action of each one of them in two BS-like operations, one for horizontally polarized light  and one for vertically polarized light, i.e.
adopting the notation introduced in~(\ref{DDD}), we describe them in terms of the following unitary transformation 
\begin{equation}
|H\rangle\langle H|\otimes BS_{H}+|V\rangle\langle V|\otimes BS_{V}\;.
\end{equation}
Here $BS_{H}$ and $BS_{V}$ are operators  coupling the vectors $|\mbox{\bf a}\rangle$, $|\mbox{\bf b}\rangle$  as in~(\ref{MATR}) with normalized transmissivities  $T'_{H}$, $T'_{V}$ and reflectivities $R'_{H}$, $R'_{V}$,  connected with the corresponding optical values  $T_{H}$, $T_{V}$, $R_{H}$, and $R_{V}$ via the associated losses $L_H$ and $L_V$. 
These parameters have been characterized sperimentally by using a CW diode laser with the same wavelength ($810nm$) of the expected entangled photons, obtaining
on average over the set of the PBSs employed in the experiment, the values 
 $T_{H}=0.965$, $R_{H}=0.0185$ and $L_{H}=0.022$ ($T_{V}=0.004$, $R_{V}=0.948$ and $L_{V}=0.048$). 

\begin{figure}[t]
\includegraphics[width=0.96 \columnwidth]{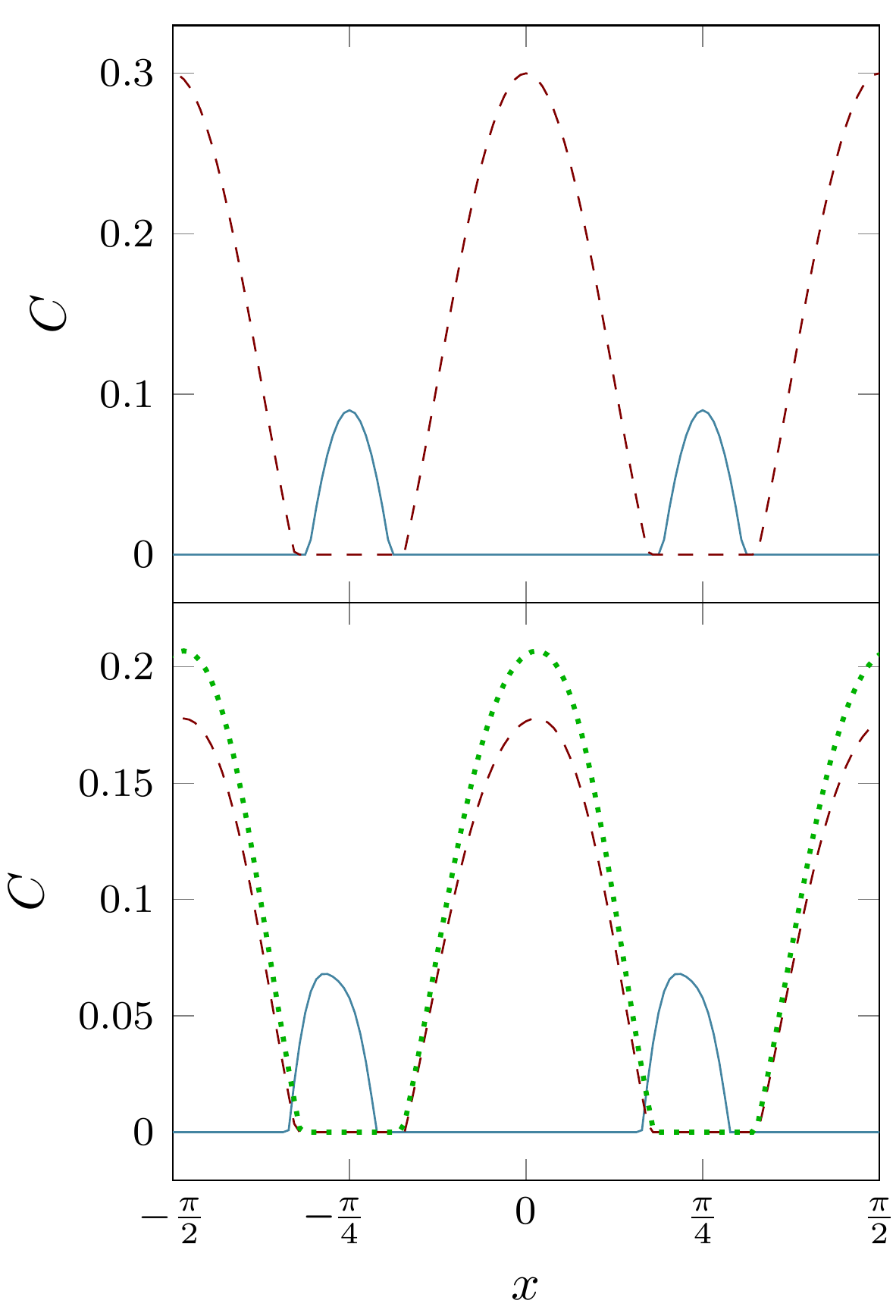}
\caption{\textbf{Top:} Concurrence  values at the output of the scheme of Fig.~\ref{fig:Scheme}  obtained assuming ideal conditions. The blue curve gives $C$ for the map $\mathcal{M'}=\Phi\circ\Psi\circ\Phi\circ\Psi$ of 
Eq.~(\ref{sequence})  obtained by rotating the angle $x=\theta$ with $\varphi$ fixed at $\frac{\pi}{4}$;  the red-dashed curve represents instead the value of $C$ for the channels  $\mathcal{M}_{1}=\Phi\circ \Phi$ or $\mathcal{M}_{2}=\Psi\circ \Psi$, obtained by rotating the angle $x=\theta$ or $x=\varphi$ respectively. \textbf{Bottom:} Same plots as the Top section obtained by using real optical-elements simulations. Also in this case the blue curve represents $\mathcal{M'}$ obtained by rotating the angle $x=\theta$ with $\varphi$ fixed at $\frac{\pi}{4}$; the red-dashed curve represents instead $\mathcal{M}_{1}$, obtained by rotating the angle $x=\theta$ while the the green-dotted curve represents $\mathcal{M}_{2}$, obtained by rotating the angle $x=\varphi$. All the curves have been produced by taking the transmission coefficients of the active AD channels equal to $\eta=0.3$.}
\label{fig:simulations}
\end{figure}

In Fig.~\ref{fig:simulations} we report a comparison between the ideal theoretical behaviour of the output concurrences $C$ and the corresponding simulated values obtained by rescaling the setup parameters as detailed above. 
Apart from exhibiting degraded values of $C$, one notices that the simulated values present a clear difference in the behaviour of  ${\cal M}_1$ and  ${\cal M}_2$  and a crooked 
 asymmetry in the concurrence peaks of ${\cal M}'$  which are not present in the ideal theoretical curves but which 
are clearly evident in the experimental data of Fig.~\ref{fig:Cut_and_pasteNEW}.
The residual  disagreement between the latter and the simulations   must be attributed to the different fiber coupling efficiencies of all the 27 possible polarization-path modes which are
difficult to chart due to mechanical random fluctuations of the setting. 

\end{document}